\documentclass[aps,prb,twocolumn,superscriptaddress]{revtex4}
\usepackage{graphicx}
\usepackage{amssymb}
\usepackage{amsmath}
\usepackage{graphicx}
\usepackage{epsfig}
\usepackage{color}
\usepackage{bm}

\begin{document}

\title{Spin current Kondo effect in frustrated Kondo systems}
\author{Jiangfan Wang}
\affiliation{Beijing National Laboratory for Condensed Matter Physics, Institute of Physics,
Chinese Academy of Science, Beijing 100190, China}
\author{Yi-feng Yang}
\email[]{yifeng@iphy.ac.cn}
\affiliation{Beijing National Laboratory for Condensed Matter Physics,  Institute of Physics, 
Chinese Academy of Science, Beijing 100190, China}
\affiliation{School of Physical Sciences, University of Chinese Academy of Sciences, Beijing 100190, China}
\affiliation{Songshan Lake Materials Laboratory, Dongguan, Guangdong 523808, China}
\date{\today}

\begin{abstract}
Magnetic frustrations can enhance quantum zero-point motion in spin systems and lead to exotic topological insulating states. When coupled to mobile electrons, they may lead to unusual non-Fermi liquid or metallic spin liquid states whose nature has not been well explored. Here, we propose a spin current Kondo mechanism underlying a series of non-Fermi liquid phases on the border of Kondo and magnetic phases in a frustrated three-impurity Kondo model. This mechanism is confirmed by renormalization group analysis and describes movable Kondo singlets called ``holons'' induced by an effective coupling between the spin current of conduction electrons and the vector chirality of localized spins. Similar mechanism may widely exist in all frustrated Kondo systems and be detected through spin-current noise measurements.
\end{abstract}

\maketitle

\section{Introduction}\label{section1}
Kondo interactions between localized spins and mobile electrons underly many exotic quantum many-body states in condensed matter physics \cite{review_si2010,YangPNAS2017}. Geometric frustration adds additional richness to this problem \cite{coleman2010frustration}. In frustrated Kondo lattice systems, experiments have uncovered rich phase diagrams with unusual non-Fermi liquid (NFL) or metallic spin liquid phases \cite{zhao2019CePdAl,Canfield2011YbAgGe,Nakatsuji2006PrIrO, Kim2013YbPtPb}. The nature of these exotic phases is still under debate \cite{coleman2010frustration,zhao2019CePdAl,Wang2020}. A minimum model for studying the interplay between geometric frustrations and the many-body Kondo coupling   is the three impurity Kondo model (3IKM) \cite{Ferrero2007,Konig2020,Coleman2021Hund,Kudasov2002,Paul1996,Ingersent2005,Lazarovits2005}. 

Geometric frustrations typically lead to static or fluctuating noncolinear or noncoplanar configurations measured by the spin vector chirality ($\bm{S}_i\times \bm{S}_j$) or scalar chirality ($\bm{S}_i\times \bm{S}_j )\cdot \bm{S}_k$ \cite{Wen1988chiral,Grohol2005,McCulloch2008,Machida2010PrIrO}. When coupled to mobile electrons, the spin chirality may give rise to anomalous electron transport properties restricted by parity ($\mathcal{P}$) and time-reversal ($\mathcal{T}$) symmetries. For example, a classical spin trimer with nonzero vector chirality may cause spin Hall effect \cite{Ishizuka2019}, while that with nonzero scalar chirality can induce anomalous Hall effect \cite{Ishizuka2019,Ishizuka2018}. Similarly, in a Mott insulator \cite{Batista2008} or a nanoscale conducting ring coupled to ferromagnetic leads \cite{Tatara2003}, scalar chirality can induce circular electric currents. The correlation between spin vector chirality and electron spin current may be enhanced by the Kondo effect \cite{YDWang2020}. But how their strong coupling may affect the many-body ground state has not been explored. 

In this work, we study the 3IKM with a dynamical large-$N$ Schwinger boson approach \cite{Yashar1D,Komijani2019,wang2019quantum,Wang2020}, combined with renormalization group (RG) analysis. We found that the spin vector chirality can induce an effective spin current Kondo effect (SCK) and  causes a series of intermediate NFL phases featured with a fractional holon phase shift  ($\pi/3$ or $2\pi/3$) due to partial Kondo screening. These phases emerge on the border of magnetic and Kondo phases and may be viewed as a quantum superposition of local Kondo singlets on different impurity sites. Evidently, a prerequisite of the SCK is the existence of electron spin or charge currents. Previous studies on 3IKM usually assumed independent electron baths \cite{Ferrero2007,Konig2020,Coleman2021Hund} and thus forbid such current flow. Our work is based on an improved treatment of the nonlocal spin current interaction in a shared-bath model and the NFL ground states may have a lower energy due to the transfer of Kondo singlets between impurity sites  \cite{Kudasov2002}. In fact, numerical RG analyses have previously discovered stable NFL fixed points within the shared-bath 3IKM \cite{Paul1996,Ingersent2005,Lazarovits2005} but not in the independent-bath model when $C_3$ symmetry is present \cite{Ferrero2007}. The SCK mechanism may also be responsible for the metallic spin liquid in the frustrated Kondo lattice.

\section{Model and Methods}\label{sec:2}
\begin{figure}[t]
\centering\includegraphics[scale=0.265]{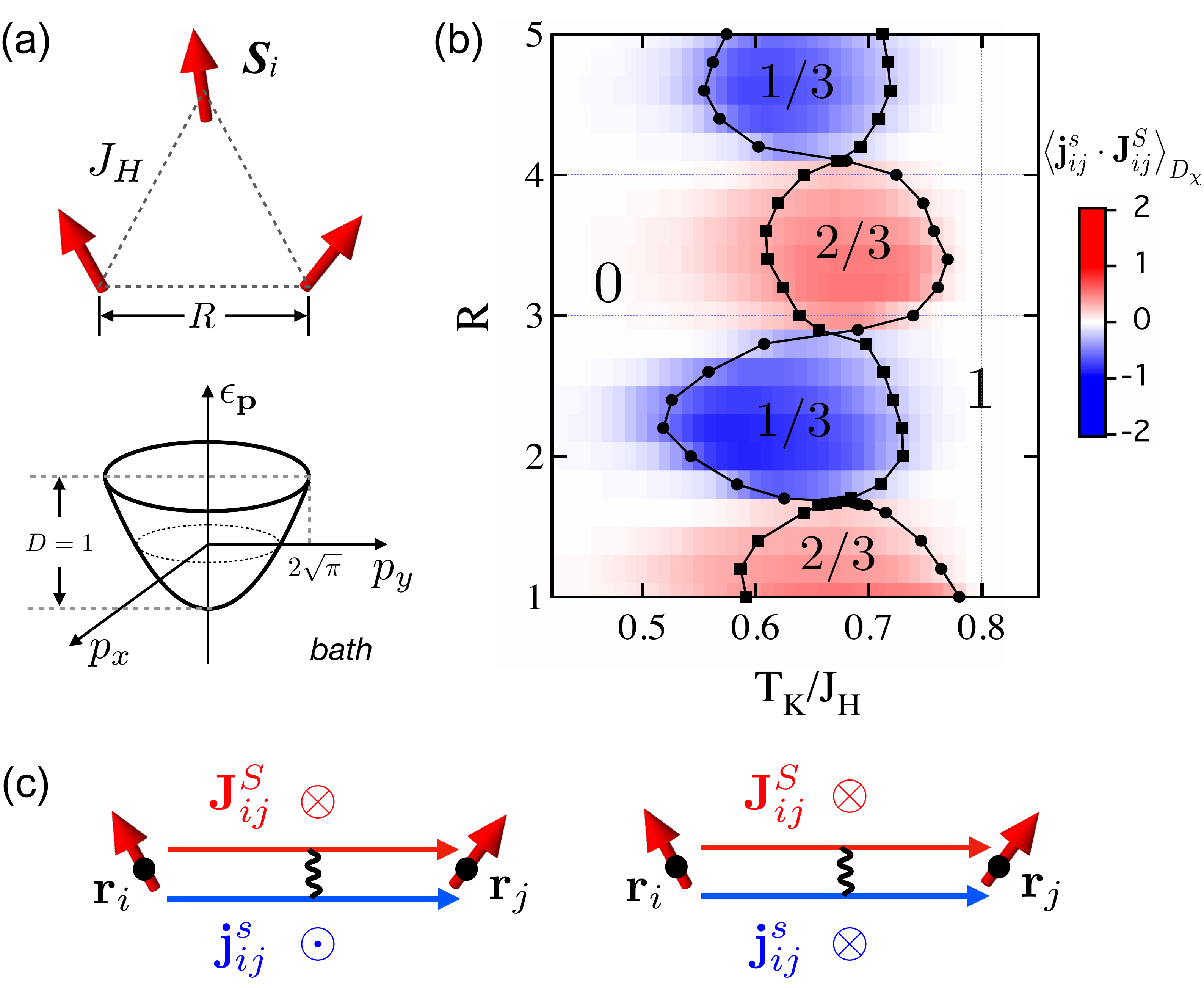}
\caption{(Color online) The three-impurity Kondo model (3IKM) and its phase diagram. (a) Schematic picture of 3IKM, showing three antiferromagnetically coupled impurity spins (red bold arrows) sitting on the vertices of an equilateral triangle with side length $R$ and Kondo coupled to a two-dimensional electron bath with a parabolic dispersion. (b) The large-$N$ phase diagram of 3IKM in terms of $R$ and $T_K/J_H$. The black curves with data points are the phase boundaries determined by the jump of holon phase shift, $\delta_\chi$. The number in each phase denotes the value $\delta_\chi/\pi$. The intensity plot with a red-blue color scale shows the correlation between the conduction electron spin current $\mathbf{j}_{ij}^s$ and the impurity spin vector chirality $\mathbf{J}_{ij}^S\equiv \mathbf{S}_i\times \mathbf{S}_j$. (c) Schematic pictures illustrating the antiferromagnetic (left) and ferromagnetic (right) spin current correlations. The symbol $\otimes$ and $\odot$ denote two opposite spin polarizations of the spin current. }
\label{fig:phase}
\end{figure}

We start with the following Hamiltonian of 3IKM:
\begin{equation}
H=\sum_{{\bm p}a\alpha}\epsilon_{\bm{p}} c_{{\bm p}a\alpha}^\dagger c_{{\bm p}a\alpha}+J_K\sum_{i=1}^3\bm{s}_{i}\cdot \bm{S}_{i}+J_H\sum_{\left\langle ij\right\rangle}\bm{S}_{i}\cdot \bm{S}_{j},
\label{H}
\end{equation}
where $a$ and $\alpha$ denote the channel (orbital) and spin indices of conduction electrons, $i$ labels the three vertices of an equilateral triangle with side length $|{\bm{r}}_i-{\bm{r}}_j|=R$ (see Figure \ref{fig:phase}(a)). The electron dispersion is chosen as $\epsilon_{\bm p}={\bm p}^2/2\pi-1$, with $|\bm{p}|\leq 2\sqrt{\pi}$ and half-band width $D=1$, to ensure the $C_3$ symmetry of the Hamiltonian and simplify the calculations.  $\bm{s}_i=\frac{1}{2}\sum_{a\alpha\beta}c_{ia\alpha}^\dagger \bm{\sigma}_{\alpha\beta}c_{ia\beta}$ is the electron spin density at ${\bm r}_i$, and $\bm{S}_i$ is the local spin. The Heisenberg term describes antiferromagnetic exchange interaction between local spins and is assumed to be independent of $R$.

The Schwinger boson representation has been widely used in studying frustrated spin systems \cite{Read1991SpN,Wang2006PSG,FlintSpN2009}.  In this representation, the local spin operator is written as $\bm{S}_i=\frac{1}{2}\sum_{\alpha\beta}b_{i\alpha}^\dagger \bm{\sigma}_{\alpha\beta}b_{i\beta}$, with a constraint fixing the number of bosons at each site, $n_b(i)=2S$, where $S$ is the spin size. Using a mean-field like decomposition, the Heisenberg term becomes
\begin{equation}
J_H\bm{S}_i\cdot \bm{S}_{j} \rightarrow \left(\tilde{\Delta}_{ij}^\dagger \Delta_{ij}-\tilde{\Gamma}_{ij}^\dagger \Gamma_{ij}\right)+H.c.+\frac{4(|\Delta_{ij}|^2-|\Gamma_{ij}|^2)}{J_H},
\label{HS-2}
\end{equation}
where $\tilde{\Delta}_{ij}^\dagger \equiv \sum_{\alpha}\text{sgn}(\alpha)b_{i\alpha}^\dagger b_{j,-\alpha}^\dagger $ creates a singlet valence bond, and  $\tilde{\Gamma}_{ij}\equiv \sum_{\alpha}b_{i\alpha}^\dagger b_{j\alpha}$ is the boson hopping term. $\Delta_{ij}$ and $\Gamma_{ij}$ are two auxiliary fields in the functional integral, with mean-field values $\Delta_{ij}=-J_H\left\langle \tilde{\Delta}_{ij}\right\rangle /4$ and $\Gamma_{ij}=-J_H\left\langle \tilde{\Gamma}_{ij}\right\rangle /4$. In the mean-field theory, the constraint is enforced on average by adding a lagrange multiplier term $\sum_i\lambda_i(n_b(i)-2S)$ to the Hamiltonian. Without Kondo coupling ($J_K=0$), the ground state of the Heisenberg model has a total effective spin $S_{tot}=1/2$ for half-integer spins and $S_{tot}=0$ for integer spins. The former becomes Kondo screened immediately by turning on a small $J_K$, while the latter remains unscreened within a finite range of $J_K$.

In the limit of large $J_K$, the physics is dominated by the Kondo term, which can be decoupled by introducing a spinless fermionic holon field $\chi_{ia}$ that creates a Kondo quasi-bound state \cite{Wang2020},
\begin{equation}
J_K\bm{s}_i\cdot \bm{S}_i\rightarrow \sum_{a\alpha}\frac{1}{\sqrt{2}}b_{i\alpha}^\dagger c_{ia\alpha}\chi_{ia}+H.c.+\sum_a \frac{\chi_{ia}^\dagger \chi_{ia}}{J_K}.
\label{HS-1}
\end{equation}
The effective onsite energy of holon decreases as one increases $J_K$, and becomes negative at large $J_K$.  It is then energetically favorable to bind a spinon and a conduction hole into a holon, forming a Kondo singlet below the characteristic Kondo temperature $T_K=De^{-2D/J_K}$. For $T_K/J_H\gg 1$, all impurities are Kondo screened, so that each impurity site is  occupied by a holon with positive electric charge. The charge conservation ensures equal numbers of negative charges being released into the electron Fermi sea, leading to a Kondo resonance peak in the Kondo impurity model, or a large electron Fermi surface in the lattice model \cite{Coleman2005sum, Wang2020}.

To study the physics at intermediate $T_K/J_H$, we perform a large-$N$ calculation, with $\alpha=\pm1,\cdots, \pm N/2$, $a=1,\cdots, K$, and $\kappa\equiv K/N=2S/N$ fixed ($\kappa=0.2$ in this paper). Here, we focus on the uniform real solution, $\Delta_{ij}=\Delta$, $\Gamma_{ij}=\Gamma$ and $\lambda_i=\lambda$. The $C_3$ symmetry allows us to take Fourier transform along the triangle, $b_{j\alpha}=\frac{1}{\sqrt{3}}\sum_h b_{h\alpha}e^{i2\pi h j/3}$ (same for $\chi$ and $c$ fields), where $h=0,\pm 1$ is the ``helicity'' number. The large-$N$ solution is obtained by solving the following self-consistent equations  (see Appendix A1):
\begin{subequations}
\begin{align}
\Sigma _{b}(h, i\nu_n)=&-\frac{\kappa}{3\beta} \sum_{h'l}G_{\chi }(h-h', i\nu_n -i\omega_l)g_c(h', i\omega_l),  \\
\Sigma _{\chi}(h, i\omega_n)=&\frac{\kappa}{3\beta} \sum_{h'l}G_{b }(h+h', i\omega_n +i\omega_l)g_c(h', i\omega_l), 
\end{align}
\label{eq:SelfE}
\end{subequations}
where
\begin{subequations}
\begin{align}
g_c(h,i\omega_n)=&\frac{1}{\mathcal{V}}\sum_{\bm p}\frac{1+\frac{2}{3}\sum_{j}\cos \left(\frac{2\pi h}{3}+\bm{p}\cdot (\bm{r}_{j}-\bm{r}_{j+1})\right) }{i\omega_n-\epsilon_{\bm p}},  \\
G_b(h,i\nu_n)=&\frac{1}{\gamma_b(h,i\nu_n)-3\Delta^2\delta_{|h|,1}/\gamma_b(-h,-i\nu_n)}, \\
G_\chi(h,i\omega_n)=&\frac{1}{-1/J_K-\Sigma_\chi(h,i\omega_n)},
\end{align}
\label{eq:GF}
\end{subequations}
are the Green's functions of conduction electrons, spinons and holons,  respectively. Here $\gamma_b(h,i\nu_n)=i\nu_n-\lambda+2\Gamma \cos (2\pi h/3)-\Sigma_b(h,i\nu_n)$,  and $i\nu_n$ ($i\omega_n$) is the bosonic (fermionic) Matsubara frequency. $\Delta$, $\Gamma$ and $\lambda$ are treated as variational parameters. The Green's functions are symmetric under inversion of helicity $h\rightarrow -h$, which is ensured by the real $\Gamma$ and the symmetric dispersion $\epsilon_{\bm{p}}$. In the limit of $R=|\bm{r}_i-\bm{r}_j|\rightarrow \infty$, both $g_c$ and $G_\chi$ become independent of $h$, and our model reduces to the independent-bath model.  In the limit $R=0$, one has $g_c(h,\omega)=g_c(0,\omega)\delta_{h0}$ and only the $h=0$ electron states are coupled to the impurities. In this paper, we focus on the intermediate range $1\leq R\leq 5$ that is physically relevant for real materials.

\section{Results}\label{sec:3}

\subsection{Phase diagram}\label{sec:3.1}

The zero temperature phase diagram in terms of $T_K/J_H$ and $R$  is shown in Figure \ref{fig:phase}(b), containing a series of intermediate regions determined by the jump of the holon phase shift $\delta_\chi$. At small $T_K/J_H$ and for integer spins, the three local spins form a total singlet with $\delta_\chi/\pi=0$.  At large $T_K/J_H$, the spins are fully Kondo screened, with a maximal phase shift $\delta_\chi/\pi=1$. At intermediate $T_K/J_H$, two different phases with $\delta_\chi/\pi=1/3$ and $2/3$ occur alternately with increasing $R$. The non-integer $\delta_\chi/\pi$ indicates that one or two of the spins are Kondo screened on average.

\begin{figure}[t]
\centering
\includegraphics[scale=0.4]{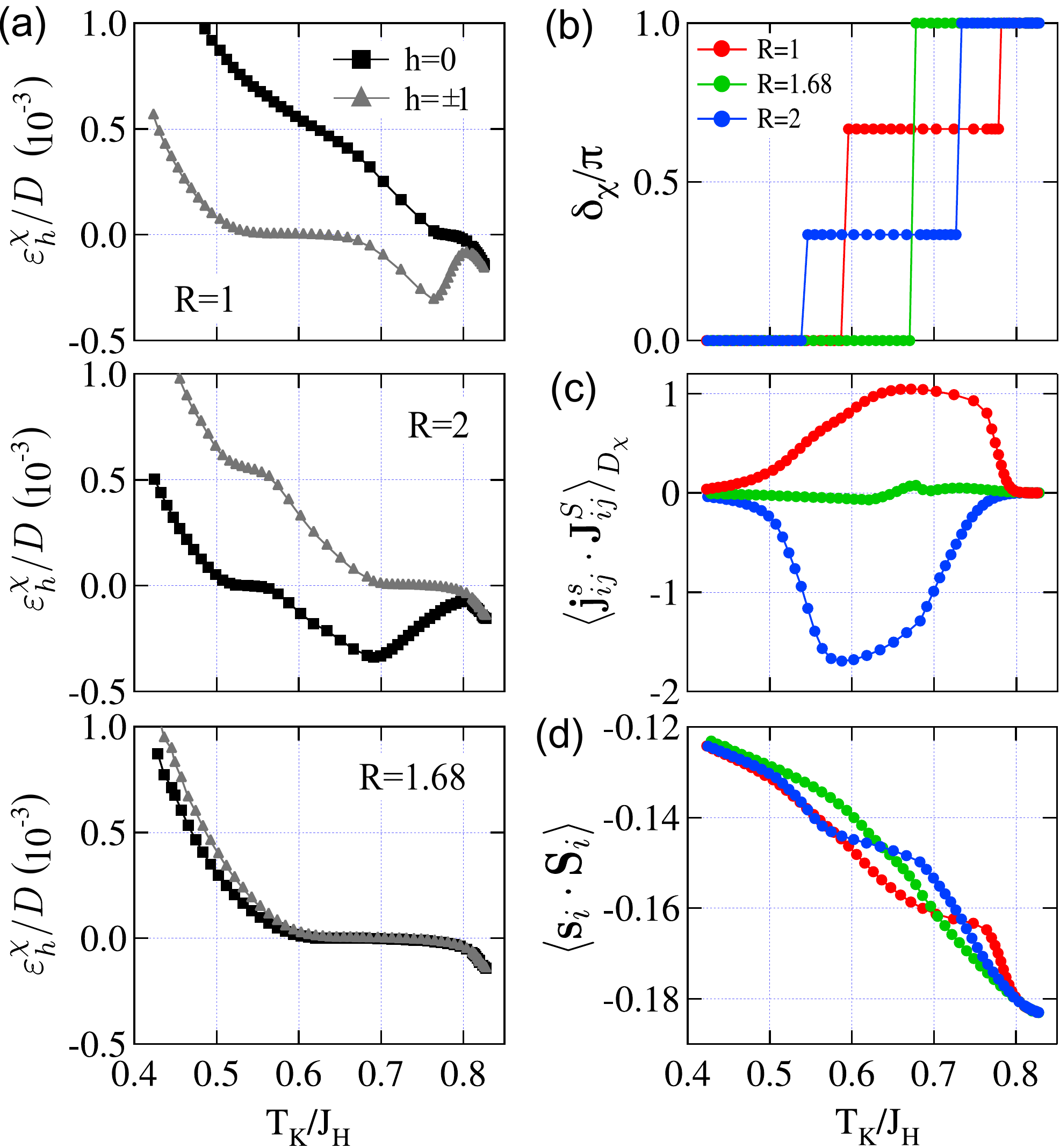}
\caption{(Color online) Holon phase shift and spin current correlations. (a) Evolution of the holon energy levels with  $T_K/J_H$ for $R=1$, $2$ and $1.68$. (b-d) The holon phase shift, spin current correlation, and local Kondo correlation as functions of $T_K/J_H$. Data for $R=1$, $2$ and $1.68$ are shown with red, green and blue points, respectively. }
\label{fig:holon}
\end{figure}

The holon phase shift is defined as $\delta_\chi \equiv \sum_h \text{Im}\ln (-G_\chi^{-1}(h,0))/3$. At zero temperature, the quantity $3\delta_\chi/\pi=\sum_{h}\theta(-\varepsilon_h^\chi)$ ($\theta(x)$ is the step function) simply counts the number of holon levels below the Fermi energy. The holon energy level, $\varepsilon_h^\chi$, is determined by the equation $1/J_K+\Sigma_\chi'(h,\varepsilon_h^\chi)=0$. Figure \ref{fig:holon}(a) shows its evolution with $T_K/J_H$ at three different $R$. $\varepsilon_h^\chi$ evolves from above the Fermi level at small $T_K/J_H$, to  below the Fermi level at large $T_K/J_H$. For $R=1, 2$, $\varepsilon_0^\chi$ and $\varepsilon_{\pm 1}^\chi$ are well separated, leaving a finite range of $T_K/J_H$ where $\varepsilon_0^\chi$ and $\varepsilon_{\pm 1}^\chi$ are on the opposite sides of the Fermi level. However for $R=1.68$, $\varepsilon_0^\chi$ and $\varepsilon_{\pm 1}^\chi$ are nearly degenerate and cross the Fermi level at the same $T_K/J_H$.

Therefore, $\delta_\chi/\pi$ consists of several plateaus ($0, 1/3, 2/3, 1$) separated by jumps at the phase boundaries, as shown in Figure \ref{fig:holon}(b). Due to the global U(1) symmetry, there is a Ward identity, $\sum_{h}\delta_c(h)=3\delta_\chi/N$, relating the electron phase shift, $\delta_c(h)\equiv -\text{Im}\ln [1-g_c(h,0^+)\Sigma_c(h,0^+)]$, to the holon phase shift \cite{Coleman2005sum}.  Based on the Friedel sum rule, there will be a Kondo resonance peak containing $\Delta n_c=\sum_{h\alpha}\delta_c(h)/\pi=3\delta_\chi/\pi$ additional electrons per channel, indicating $\Delta n_c$ local spins being ``delocalized''. 

The holon levels seem to be ``pinned'' at the Fermi energy at the phase boundaries, consistent with previous studies on the Kondo lattice \cite{Komijani2019,wang2019quantum}. Here, well defined holon levels (with sharp quasiparticle peaks)  only appear inside the energy range $|\omega|<D_\chi\sim 10^{-3}D$ around the Fermi level, consistent with the single ion Kondo temperature $T_K/D=e^{-2D/J_K}\approx 0.0067$ that we fixed in our calculations.

\subsection{Spin current Kondo effect}\label{sec:3.2}

The intermediate phases are featured with an enhanced spin current correlation, $\left\langle \mathbf{j}_{ij}^s\cdot \mathbf{J}_{ij}^S \right\rangle_{D_\chi}$, as shown by the color-coded intensity plot of Figure \ref{fig:phase}(b). The correlation has a negative (positive) sign in the $\delta_\chi/\pi=1/3$ ($2/3$) phase, corresponding to an AFM (FM) SCK effect as illustrated in Figure \ref{fig:phase}(c). Here the electron spin current is defined as, $\bm{j}_{ij}^s=ic_{j\alpha}^\dagger \bm{\sigma}_{\alpha\beta} c_{i\beta}/2+h.c.$, and the impurity spin current is $\bm{J}_{ij}^S\equiv \bm{S}_i\times \bm{S}_j$. The identification of vector chirality as a spin current is supported by the equation of motion of the Heisenberg model, $d\bm{S}_j/dt=J_H\sum_i  \bm{S}_i\times \bm{S}_j$ \cite{Bruno2005,Katsura2005}. The subscript ``$D_\chi$'' denotes an average over the coherent part of the holon spectrum that lies within the range $|\omega| \leq D_\chi$ around the Fermi level. Physically, this accounts for the retardation of the SCK effect due to the ``slowly-moving'' holons.

The SCK effect arises because only $\varepsilon_{0}^\chi$ or $\varepsilon_{\pm 1}^\chi$ are occupied in the intermediate phases. This means that a holon can be thermally excited to a different level, producing a net ``moment'' (helicity) along the triangle. These moving holons (Kondo singlets) can mediate a nonlocal Kondo scattering process described by $c_{ja\alpha}^\dagger c_{ia\beta} b_{i\beta}^\dagger  b_{j\alpha}$ \cite{Wang2020}, which is directly related to the spin current correlation via the relation $i\bm{\sigma}_{\alpha\beta}\cdot (\bm{S}_i \times \bm{S}_j)=( b_{j\beta}^\dagger \tilde{\Gamma}_{ij}b_{i\alpha}-b_{i\beta}^\dagger \tilde{\Gamma}_{ji}b_{j\alpha})/2$.

More details on the evolution of $\left\langle \mathbf{j}_{ij}^s\cdot \mathbf{J}_{ij}^S \right\rangle_{D_\chi}$ with $T_K/J_H$ are shown in Figure. \ref{fig:holon}(c) for different $R$. We see it is substantially enhanced at intermediate $T_K/J_H$ for $R=1, 2$, but with opposite signs, and strongly suppressed for $R=1.68$. As discussed in Appendix A2, the main contribution to the spin current correlation is proportional to $n_{\pm 1}^\chi  -n_0^\chi$, where $n_h^\chi$ is the holon ``occupation number''. For $n_0^\chi>0$ and $n_{\pm 1}^\chi=0$, one has $\delta_\chi/\pi=1/3$ and the AFM SCK effect, while for $n_0^\chi=0$ and $n_{\pm 1}^\chi>0$, one has $\delta_\chi/\pi=2/3$ and the FM SCK effect. When both $n_0^\chi$ and $n_{\pm 1}^\chi$ are zero ($\delta_\chi/\pi=0$) or finite ($\delta_\chi/\pi=1$), the SCK effect is suppressed. In fact, the Pauli exclusion principle forbids holons to move when all the holon levels are occupied, or equivalently, when all the spins are fully Kondo screened. This is exactly what happens in the independent-bath model, where $n_0^\chi=n_{\pm 1}^\chi$ always holds. 

The local Kondo correlation can be calculated in a similar way, leading to $\left\langle \mathbf{s}_i\cdot \mathbf{S}_i \right\rangle =C -\frac{1}{3J_K^2}(n_0^\chi+n_1^\chi+n_{-1}^\chi)$, where $C$ represents contributions from incoherent part of the holon spectrum. As shown in Figure \ref{fig:holon}(d), the Kondo correlation is always negative (antiferromagnetic) with an absolute value increasing monotonically with $T_K/J_H$.

\subsection{Spin correlation}\label{sec:3.3}

\begin{figure}[t]
\centering\includegraphics[scale=0.44]{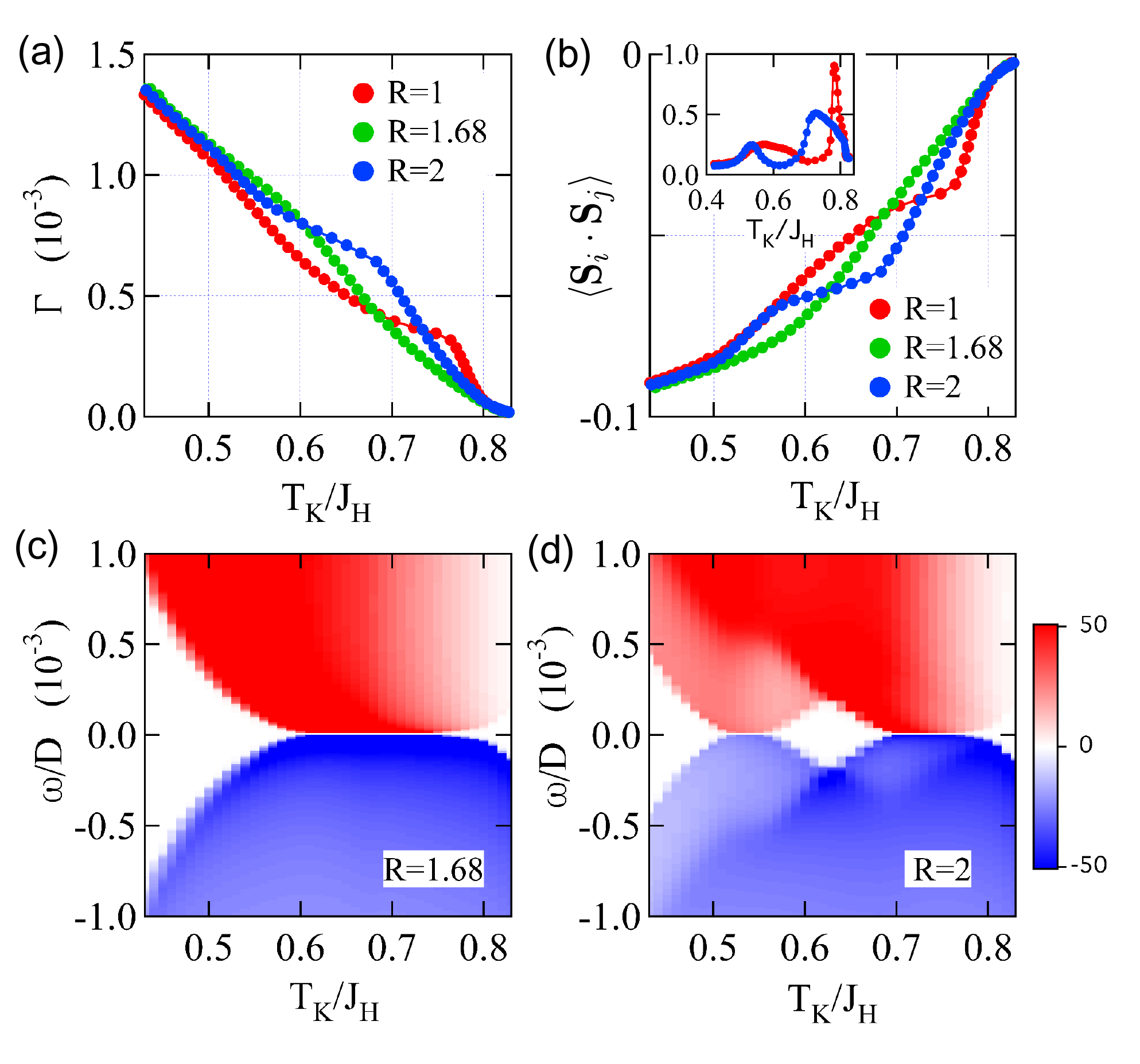}
\caption{(Color online) Spin correlations and spinon density of states. (a) The spinon hopping amplitude  as a function of $T_K/J_H$ for $R=1$, $2$ and $1.68$. 
(b) The spin correlation function $\left\langle \bm{S}_i\cdot \bm{S}_j\right\rangle$ as a function of $T_K/J_H$ for $R=1$, $2$ and $1.68$. The inset shows its derivative with respect to $T_K/J_H$ for $R=1$ and $2$.  (c,d) Intensity plots of the spinon's density of states, $-\sum_h G_b''(h,\omega)/(3\pi)$,  for $R=1.68$ and $R=2$,  showing their evolutions with $T_K/J_H$. }
\label{fig:corr}
\end{figure}

The SCK effect is a combined effect of the Kondo and inter-site magnetic correlations. It requires not only movable holons, but also movable spinons. The spin current correlation is proportional to the spinon hopping amplitude $\Gamma$. As shown in Figure \ref{fig:corr}(a), $\Gamma$ decreases monotonically with increasing $T_K/J_H$, and vanishes deep inside the Kondo phase. This is another reason why SCK effect is suppressed for large $T_K/J_H$. The spinon pairing amplitude $\Delta$ has very similar behaviors to $\Gamma$. They are related to the impurity spin correlation through  $\left\langle \bm{S}_i\cdot \bm{S}_{i+1}\right\rangle=(\Gamma^2-\Delta^2)/(4J_H^2)$. As shown in Figure \ref{fig:corr}(b), it is always negative with an absolute value decreasing monotonically with increasing $T_K/J_H$. Direct comparison of Figure \ref{fig:holon}(d) and Figure \ref{fig:corr}(b) reveals complementary behavior of the Kondo and spin correlations, showing strong competition between the Kondo effect and the magnetic interaction. Both stay relatively ``constant'' within each phase but change rapidly near the phase boundaries, as is seen from the peaks of the derivative of $\left\langle \bm{S}_i\cdot \bm{S}_{i+1}\right\rangle$ with respect to $T_K/J_H$ (inset of Figure \ref{fig:corr}(b)).

The boundary of the intermediate phases is marked with gapless holon and spinon excitations. This is seen from the intensity plots of spinon's density of states (DOS), $-\sum_h G_b''(h,\omega)/(3\pi)$, shown in Figure \ref{fig:corr}(c) and Figure \ref{fig:corr}(d) for $R=1.68$ and $2$, respectively. The gapless fractional excitations are expected to cause quantum critical behaviors \cite{Komijani2019,Wang2020}. Away from the phase boundaries, both spinon and holon develop gaps in their spectra.  Here, the intermediate phase has a small holon gap due to the discrete nature of the holon levels, which is expected to be closed in the lattice version causing a dispersive holon band and a hidden holon Fermi surface \cite{Wang2020}.

\subsection{Renormalization group analysis}\label{sec:3.4}

To clarify the origin of the SCK, we perform a simplified renormalization group (RG) analysis and integrate out the scattering processes involving high energy electrons or holes \cite{Anderson1970}. This gives a low temperature effective Hamiltonian with an emergent ``strongly coupled'' SCK term, suggesting that the SCK effect is energetically favored. 

The renormalized Hamiltonian with a reduced electron band width $D'=D-\delta D$ has a modified spin-electron scattering term $H_K'=H_K+\Delta H_K$, with
\begin{align} 
\Delta H_K=&-\frac{J_K^2\rho\delta D  }{4D}\sum_{ij}c_{ia\alpha}^\dagger c_{ja\beta}\sigma_{\alpha\gamma}^\mu \sigma_{\gamma\beta}^\nu \notag \\
 &\times\left(\sum_{\bm{q}^+} e^{i\bm{q}\cdot \bm{R}_{ij}}S_i^\mu S_{j}^\nu-\sum_{\bm{q}^-} e^{i\bm{q}\cdot \bm{R}_{ij}} S_{j}^\nu S_i^\mu  \right). \label{eq:PS}
\end{align}
Here $\mu, \nu$ denote the spatial components, $\bm{R}_{ij}=\bm{r}_i-\bm{r}_j$, and $\bm{q}^{\pm}$ satisfies $D'<\epsilon_{\bm{q}^+}<D$, $-D<\epsilon_{\bm{q}^-}<-D'$, respectively. The summation over $i, j$ contains two parts. For $i=j$, Eq. (\ref{eq:PS}) gives the usual correction to the  Kondo term, $\frac{J_K^2\rho\delta D}{2D}\sum_{i}c_{ia\alpha}^\dagger c_{ia\beta}\bm{\sigma}_{\alpha\beta}\cdot \bm{S}_i$. For $i\neq j$, it gives rise to a nonlocal scattering term,
\begin{equation}
J_K'\sum_{\left\langle ij\right\rangle}\left[\bm{j}^s_{ij}\cdot \bm{J}_{ij}^S-K_{ij}\bm{S}_i\cdot \bm{S}_{j}\right], 
\label{eq:NL}
 \end{equation}
 where $K_{ij}=c_{ia\alpha}^\dagger c_{ja\alpha}+h.c.$ and $J_K'=\frac{J_K^2 \rho\delta D}{2D}[J_0(q_+ R)-J_0(q_- R)]$, with $J_0(x)$  the Bessel function and $q_{\pm}=\sqrt{2\pi (1\pm D)}$. The scalar and vector products of spins occur simultaneously due to the identity $\sigma_{\alpha\gamma}^\mu \sigma_{\gamma\beta}^\nu=\delta_{\alpha\beta}\delta_{\mu\nu}+i\epsilon_{\mu\nu\eta}\sigma_{\alpha\beta}^{\eta}$. Once averaged over the conduction electron sea, the second term contributes a part of the well-known Ruderman-Kittel-Kasuya-Yosida (RKKY) interaction, which competes with the Kondo effect to govern the fundamental physics of a Kondo lattice or multi-impurity Kondo system. The spin current interaction term is seen to emerge on the same level of the RKKY interaction but was often ignored in the literature. As we have seen, it is exactly this term that is responsible for the intermediate  phases.
 
For simplicity, instead of performing a complete RG study of the 3IKM, we are interested here in the flow of $J_K'$ in different regions of the phase diagram. This is done via the Schrieffer-Wolff transformation of a general Hamiltonian containing both the local and nonlocal interaction terms (see Appendix A3). The RG equation for $g_K'\equiv J_K'\rho$ reads:
\begin{equation}
\frac{dg_K'}{dl}=\beta(g_K')=\frac{f_R(l)}{2}g_K^2+2g_K g_K' +O(g_K^4),
\label{eq:RG}
\end{equation}
where $g_K\equiv J_K\rho$, $l\equiv -\ln D$ and $f_R(l)\equiv J_0(q_+(l) R)-J_0(q_-(l) R)$. Therefore, $g_K'$ strongly depends on the fixed point $g^*$ that $g_K$ flows to at low temperature. In the RKKY dominated region (the magnetic phase), $g_K$ is suppressed and $g^*=0$ at zero temperature \cite{Nejati2017}. This leads to a vanished $\beta(g_K')$ in the right hand of Eq. (\ref{eq:RG}), so that $g_K'$ becomes marginal in this phase. In the fully Kondo screened phase, $g^*=\infty $, and the perturbative RG breaks down at low energy scale. However, both the RKKY and SCK terms are expected to be suppressed due to the fully quenched local spins. Between these two extremes, one expects $g_K$ to flow to some stable (NFL) or unstable (quantum critical point) intermediate fixed point $0<g^*<\infty$. This indeed happens for the quantum critical point of the two-impurity Kondo model \cite{Mitchell2012}, the Anderson or Kondo impurity model with a pseudogapped fermion bath \cite{Ingersent1998RG}, the Bose-Fermi Kondo model \cite{Si2002BFKM}, and the stable NFL fixed point of 3IKM \cite{Paul1996,Lazarovits2005}.

\begin{figure}[t]
\centering\includegraphics[scale=0.53]{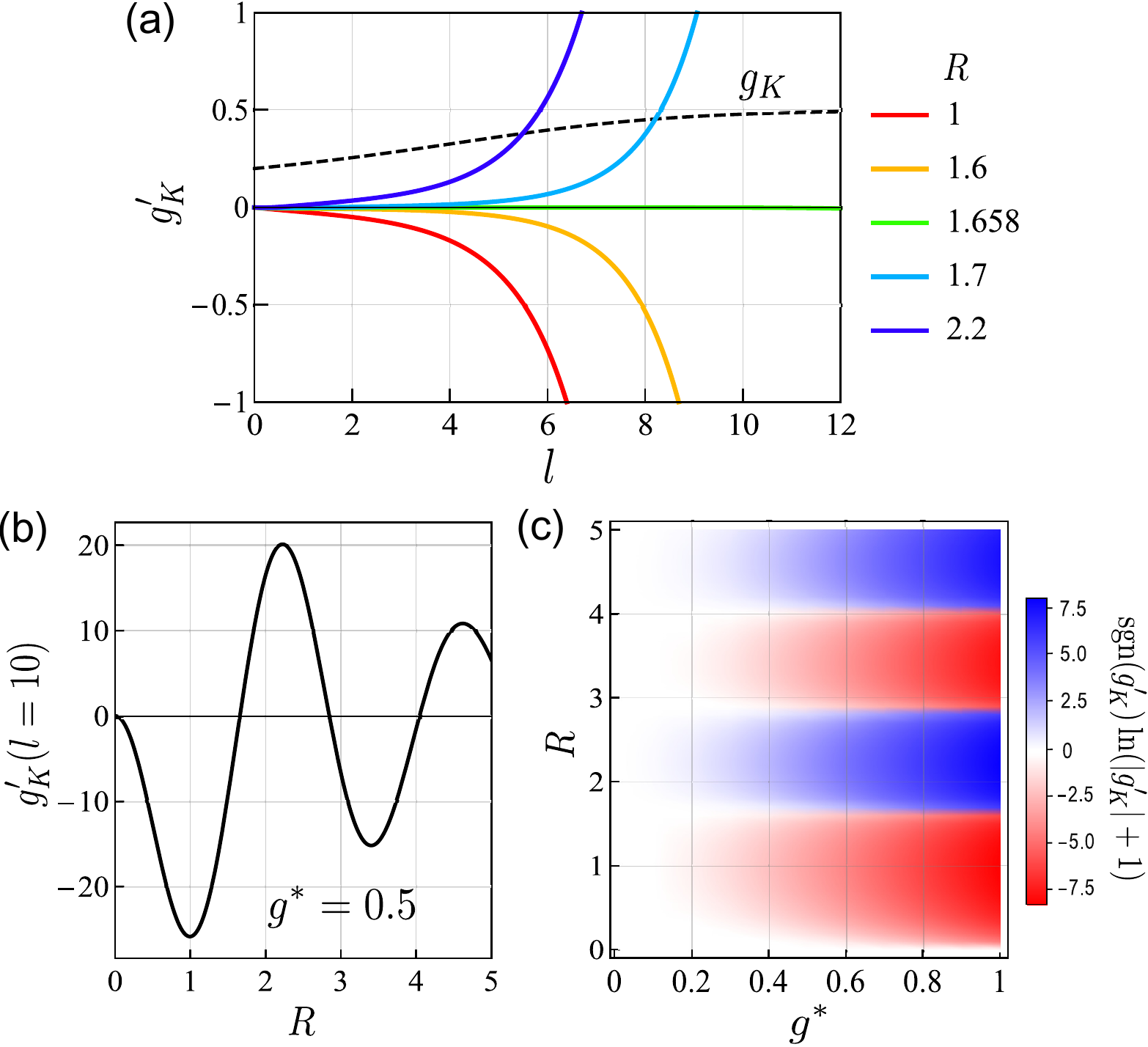}
\caption{(Color online)  Renormalization group flow.  (a) The RG flow of the coupling constant $g_K'$ for different $R$, assuming that the local Kondo coupling $g_K$ flows to a fixed point $g^*=0.5$. The initial value is $g_K=0.2$, corresponding to  $J_K=0.4$ in the large-$N$ calculations. (b) $g_K'$ as a function of $R$ at a fixed scale $l=10$ with $g^*=0.5$. (c) Intensity plot of $\text{sgn}(g_K')\ln (|g_K'|+1)$ on the $R$-$g^*$ plane at the same scale $l=10$, showing the sign change of $g_K'$ with varying $R$.}
\label{fig:RG}
\end{figure}

The behaviors of $g_K'$ may be illustrated in the intermediate coupling regime $0<g^*<1$, where the perturbative RG can be qualitatively trusted. For clarity, we assume a simplest flow equation for the Kondo coupling, $dg_K/dl=g_K^2-g_K^3/g^*$, that gives such a fixed point \cite{Nozieres1980}. Figure \ref{fig:RG}(a) compares the RG flow of $g_K$ and $g_K'$ for different $R$ with $g^*=0.5$. One finds that $g_K'$ grows rapidly to strong coupling for $R=1$ and $2.2$, but remains zero for $R\approx 1.66$. This is because the initial flow, namely the sign of $g_K'$, is controlled by the first term of Eq. (\ref{eq:RG}) that depends on $R$, while its exponential growth at large $l$ is caused by the second term of Eq. (\ref{eq:RG}). As shown in Figure \ref{fig:RG}(b), $g_K'$ at a fixed scale $l=10$ (corresponding to a small energy scale $T\sim e^{-10}D\sim 0.01 T_K$) oscillates as a function of $R$ between positive and negative, and stays zero for a series of nodes $R_n\approx 1.66, 2.84, 4.05, \cdots$. The value of $R_n$  is nearly independent of $g^*$, as can be seen from the intensity plot on the $R$-$g^*$ plane shown in Figure \ref{fig:RG}(c). As $g^*$ approaches zero, $g_K'$ also becomes negligible. The nodes $R_n$ and the associated sign change of $g_K'$ resembles qualitatively to the large-$N$ phase diagram shown in Figure \ref{fig:phase}(b). Therefore, the RG analysis supports that the strong coupling SCK effect is indeed a characteristic feature of the intermediate NFL phases and may be crucial in turning an unstable fixed point into a stable one in the 3IK model.

\section{Discussion and conclusion}\label{sec:4}

We have theoretically studied the phase diagram of 3IKM with a single shared electron bath using the dynamical large-$N$ Schwinger boson approach, focusing on the emergent NFL state and its microscopic origin. Combined with RG analysis, we show that the intermediate phase is featured with a strong SCK effect, a nonlocal Kondo scattering effect mediated by itinerant or movable Kondo singlets (holons). We noticed that Eq. (\ref{eq:NL}) can also arise in the two-impurity Kondo model (2IKM), and was demonstrated to destabilize the ``Varma-Jones'' critical point into a smooth crossover \cite{Logan2011}. In this case, applying our Schwinger boson approach does not yield an intermediate ground state \cite{Coleman-SWB-2006}. This is because that the SCK correlation is proportional to the spinon hopping amplitude $\Gamma$ which is absent in the 2IKM. Thus, frustrations are crucial for stabilizing such NFL states.

In our calculations, we have focused on the ground state solutions that keep both $\mathcal{P}$ and $\mathcal{T}$ symmetries. Due to the strong SCK coupling, it is likely that the intermediate ground states spontaneously break these symmetries, leading to nonzero $\left\langle \bm{S}_i\times \bm{S}_j \right\rangle$ and $\left\langle (\bm{S}_i\times \bm{S}_j) \cdot \bm{S}_k \right\rangle$. This can also be realized by adding a Dzyaloshinskii-Moriya interaction \cite{Moriya1960DM} or a Zeeman term to the Hamiltonian. A mean-field description of such states assumes a complex spinon hopping $\Gamma=|\Gamma|e^{i\phi/3}$, here $\phi$ is the flux through the triangle related to the scalar chirality via $\left\langle \bm{S}_1\cdot (\bm{S}_2\times \bm{S}_3)\right\rangle=\frac{N^3|\Gamma|^3}{2J_H^3}\sin (\phi)$. A persistent circular current $\left\langle J_\bigtriangleup \right\rangle=-2\text{Im}\sum_{j\alpha}\left\langle c_{j\alpha}^\dagger c_{j+1,\alpha}\right\rangle$ can be induced due to the asymmetry between $h=1$ (clockwise) and $h=-1$ (counterclockwise) electron states. In fact, an additional term $g_C J_\bigtriangleup \bm{S}_1\cdot (\bm{S}_2\times \bm{S}_3)$ with $g_C \propto g_Kg_K'$ will appear in the RG calculation due to the interference between local and nonlocal interactions. Therefore, within the intermediate phases, an applied external current will be deflected by the persistent circular current, leading to an anomalous Hall effect. Similar mechanism was studied in Ref. \cite{Ishizuka2018, Ishizuka2019} using transport theories with the assumption of classical spins. 

In reality, $J_H$ may be given by the RKKY interaction and oscillate with $R$. Then the observed phase diagram might be distorted. In particular, for a ferromagnetic spin interaction ($J_H<0$), there will be no frustration and intermediate phases \cite{Coleman2021Hund}, and the three impurity spins will align and form an effective large spin to be fully screened by conduction electrons. Nevertheless, we believe that the SCK effect is a widely existing property in frustrated Kondo systems and may be responsible for the metallic spin liquid state in frustrated Kondo lattice \cite{zhao2019CePdAl,Canfield2011YbAgGe,Nakatsuji2006PrIrO, Kim2013YbPtPb,Wang2020}. One of its consequences is the suppression or enhancement of the thermal spin-current noise, $\left\langle (\mathbf{j}^s+ \mathbf{J}^S)^2 \right\rangle=\left\langle (\mathbf{j}^s)^2 \right\rangle +\left\langle (\mathbf{J}^S)^2 \right\rangle +2 \left\langle \mathbf{j}^s\cdot \mathbf{J}^S \right\rangle$, which can be transformed into charge noise by the inverse spin Hall effect and is experimentally measurable \cite{Kamra2014SHN}. The spin-current noise should therefore be reduced (enhanced) in the presence of AFM (FM) SCK effect. Our prediction may be verified in future experiment on frustrated Kondo lattice systems.

This work was supported by the National Key R\&D Program of MOST of China (Grant No. 2017YFA0303103), the National Natural Science Foundation of China (NSFC Grant No. 12174429, No. 11774401,  No. 11974397), the Strategic Priority Research Program of the Chinese Academy of Sciences (Grant No. XDB33010100), and the Youth Innovation Promotion Association of CAS.



\begin{appendix}

\renewcommand{\thesection}{A}

\section{}

\subsection{\label{sec:level1} Action and self-consistent equations}

With the uniform real mean-field assumption, the action of 3IKM is written as:
\begin{align}
S=&S_{c0}+\int_\tau \left\{\sum_{h\alpha} b_{h\alpha}^\dagger \left( \partial_\tau+\lambda-2\Gamma \cos \frac{2\pi h}{3} \right)b_{h\alpha}-6S\beta \lambda \right. \notag \\
& +i\Delta \sum_{h\alpha}\text{sgn}(\alpha) b_{h\alpha}b_{-h,-\alpha}\sin\frac{2\pi h}{3}+c.c.  + \frac{3\beta N}{J_H}(\Delta^2-\Gamma^2)\notag \\
&+\left.\frac{1}{\sqrt{3N}}\sum_{hh'a\alpha}b_{h\alpha}^\dagger c_{h'a\alpha}\chi_{h-h',a}+c.c.+\sum_{ha}\frac{|\chi_{ha}|^2}{J_K} \right\},
\end{align}
where $S_{c0}=-\sum_{hna\alpha}c_{hna\alpha}^\dagger g_c^{-1}(h,i\omega_n)c_{hna\alpha}$ describes the free conduction electrons. The only vertex is the spinon-electron-holon three point vertex, which gives rise to the form of self-energies in Eq. (\ref{eq:SelfE}) at large-$N$. The Green's function for each field can be conveniently derived from the action. The three mean-field variables $\Delta$, $\Gamma$ and $\lambda$ are determined through the minimization equations $\partial \ln Z/\partial\Delta=\partial \ln Z/\partial\Gamma=\partial \ln Z/\partial\lambda=0$, where $Z=\int \mathcal{D}[b,c,\chi]\exp [-S]$ is the partition function. This gives
\begin{subequations}
\begin{align}
\kappa= & -\frac{1}{3\beta}\sum_{hn}G_b(h,i\nu_n)\\
\frac{\Gamma}{J_H}= & \frac{1}{3\beta}\sum_{hn}G_b(h,i\nu_n)\cos\frac{2\pi h}{3} \\
\frac{\Delta}{J_H}= & \frac{1}{3\beta}\sum_{hn}F_b(h,i\nu_n)i\sin\frac{2\pi h}{3}, 
\end{align}
\end{subequations}
where $F_b(h,i\nu_n)\equiv -\text{sgn}(\alpha)\left\langle b_{hn\alpha} b_{-h,-n,-\alpha}\right\rangle$ is the  anomalous Green's function of spinon. 

\subsection{\label{sec:level2} Spin current correlation function}

Using the SU(2) Schwinger boson representation, we can write the spin current correlation as
\begin{align}
\left\langle \mathbf{j}_{ij}^s\cdot \mathbf{J}_{ij}^S\right\rangle =&-\frac{1}{4}\sum_{a\alpha\beta\gamma} \left(\left\langle c_{ja\alpha}^\dagger c_{ia\beta}b_{i\beta}^\dagger b_{i\gamma}b_{j\gamma}^\dagger b_{j\alpha} \right\rangle \right. \notag \\
 &\left. -\left\langle c_{ja\alpha}^\dagger c_{ia\beta}b_{j\beta}^\dagger b_{j\gamma}b_{i\gamma}^\dagger b_{i\alpha} \right\rangle \right)+c.c..
\label{eq:JC}
\end{align} 
We then calculate the above expression in the large-$N$ limit. The leading order contraction of the  six-body average contains the term $\sum_\gamma\left\langle b_{i\gamma}b_{j\gamma}^\dagger \right\rangle=\sum_\gamma\left\langle b_{j\gamma}b_{i\gamma}^\dagger \right\rangle=-\frac{N\Gamma}{J_H}$. The leading order Feynman diagrams of the remaining four-body averages have order of magnitudes $\sum_{a\alpha\beta}\left\langle c_{ja\alpha}^\dagger c_{ia\beta}b_{i\beta}^\dagger b_{j\alpha} \right\rangle\sim O(NK)$ and  $\sum_{a\alpha\beta}\left\langle c_{ja\alpha}^\dagger c_{ia\beta}b_{j\beta}^\dagger b_{i\alpha} \right\rangle\sim O(K)$, respectively. The latter is sub-leading due to an additional $\delta$-function, thus can be neglected. The former is calculated as
\begin{align}
  &\sum_{a\alpha\beta}\left\langle c_{ja\alpha}^\dagger (\tau) c_{ia\beta} (\tau') b_{i\beta}^\dagger (\tau') b_{j\alpha}(\tau) \right\rangle \notag\\
=&-\frac{NK}{3\beta}\sum_{hn}\Sigma_\chi(h,i\omega_n)^2 G_\chi(h,i\omega_n)e^{i2\pi h(j-i)/3}e^{i\omega_n(\tau'-\tau)} \notag\\
=&\frac{NK}{3}\int_{-\infty}^\infty \frac{d\omega}{\pi}\frac{e^{\omega(\tau'-\tau)}}{e^{\omega/T}+1}\left[\frac{G_\chi''(0,\omega)-G_\chi''(1,\omega)}{J_K^2}\right. \notag \\
 &\left.-\Sigma_\chi''(0,\omega)+\Sigma_\chi''(1,\omega)\right].
\end{align}
Instead of equal-time correlation, here we take $\tau'-\tau\sim D_\chi^{-1}$ to account for the  time needed for the heavy holons to propagate from site $i$ to site $j$. At zero temperature $T=0$, the exponential factor effectively reduces the range of frequency integral to $-D_\chi\leq \omega \leq 0$. Practically, we calculate the following quantity:
\begin{align}
\left\langle \mathbf{j}_{ij}^s\cdot \mathbf{J}_{ij}^S\right\rangle_{D_\chi} \equiv &\frac{2\Gamma}{3J_H D_\chi}\int_{-D_\chi}^0 \frac{d\omega}{\pi}\left[\frac{G_\chi''(0,\omega)-G_\chi''(1,\omega)}{J_K^2}\right. \notag \\
  &\left. -\Sigma_\chi''(0,\omega)+\Sigma_\chi''(1,\omega)\right],
\label{eq:scorr}
\end{align}
where we have multiplied the right hand side of Eq. (\ref{eq:JC}) with a factor $4/(N^2 K D_\chi)$ to give an average-over-$D_\chi$ result with order $O(1)$ at large-$N$. We take $D_\chi=0.001$ to account for the holon spectrum near the Fermi energy, containing only well-defined quasiparticles with negligible imaginary part of self-energies. In this way, Eq. (\ref{eq:scorr}) is dominated by the term proportional to $n_1^\chi  -n_0^\chi$, where $n_h^\chi=-\int_{-D_\chi}^0 \frac{d\omega}{\pi}G_\chi''(0,\omega)$ is the holon ``occupation number''.

\subsection{\label{sec:level3} Derivation of RG equations}

To derive the RG equation of $J_K'$ to the  order $O(J_K^3)$, we start with the following general Hamiltonian:
\begin{align}
H(D)=&H_{c0}(D)+J_K\sum_i \bm{s}_i\cdot \bm{S}_i \\
+&J_K'\sum_{\left\langle ij \right\rangle}\left(\bm{j}_{ij}^s\cdot\bm{J}_{ij}^S-K_{ij}\bm{S}_i\cdot \bm{S}_j\right) \notag 
  +J_H'\sum_{\left\langle ij \right\rangle}\bm{S}_i\cdot \bm{S}_j,
\end{align}
where $H_{c0}(D)$ is the free conduction electron term, and $D$ is the half band width of conduction band. The last term arises when integrating out the high energy conduction holes and contributes a major part of the RKKY interaction, $J_{RKKY}=J_H'-J_K'\left\langle K_{ij} \right\rangle$. In the second-order perturbation, it does not enter the RG flow of $J_K'$ and is therefore dropped in the following analysis. The renormalized Hamiltonian with a reduced half band width $D'=D-\delta D$ can be obtained using the standard Schrieffer-Wolff transformation and yields
\begin{align}
H(D')=&H_{c0}(D')+J_K(D')\sum_i \bm{s}_i\cdot \bm{S}_i \notag \\
 &+J_K'(D')\sum_{\left\langle ij \right\rangle}\left(\bm{j}_{ij}^s\cdot\bm{J}_{ij}^S-K_{ij}\bm{S}_i\cdot \bm{S}_j\right) \notag \\
 &+J_C(D')J_\bigtriangleup \bm{S}_1\cdot (\bm{S}_2\times \bm{S}_3)+\cdots
\end{align}
where $``\cdots"$ contains higher order scattering terms and pure spin interaction terms. The renormalized coupling constants in the above equation read 
\begin{subequations}
\begin{align}
J_K(D')=&J_K+J_K^2\rho\delta \ln D,  \\
J_K'(D')=&J_K'+\left[2J_K J_K'+\frac{J_K^2}{2}f_R(D) \right]\rho\delta \ln D,  \\
J_C(D')=&-\frac{J_KJ_K'}{2}f_R(D) \rho\delta \ln D, 
\end{align}
\end{subequations}
where $f_R(D)\equiv J_0(q_+(D)R)-J_0(q_-(D)R)$. By defining $g_K\equiv J_K\rho$, $g_K'\equiv J_K'\rho$, $g_C\equiv J_C\rho$ and  $l\equiv -\ln D$, the above equations can be rewritten as 
\begin{subequations}
\begin{align}
\frac{\delta g_K}{\delta l}=&g_K^2,  \\
\frac{\delta g_K'}{\delta l}=&\frac{g_K^2}{2}f_R(l)+2g_K g_K',  \\
\frac{\delta g_C}{\delta l}=&\frac{g_K g_K'}{2}f_R(l).
\end{align}
\end{subequations}
The Schrieffer-Wolff transformation only gives an order $O(J_K^2)$ beta function for $g_K$. However, the beta functions for $g_K'$ and $g_C$ are of order $O(J_K^3)$, since $g_K' \sim O(J_K^2)$. At this order,  the coupling constant $g_C$ does not enter the other RG equations, and is completely determined by the flows of $g_K$ and $g_K'$.

\end{appendix}

\end{document}